\documentclass{jetpl}
\usepackage{amssymb}
\usepackage{graphicx}

\twocolumn
\lat

\begin{document}

\title{Normal state resistivity of Ba$_{1-x}$K$_x$Fe$_2$As$_2$: evidence for
multiband strong-coupling behavior}
\rtitle{Normal state resistivity of Ba$_{1-x}$K$_x$Fe$_2$As$_2$: evidence for
multiband strong-coupling behavior}
\sodtitle{Normal state resistivity of
Ba$_{1-x}$K$_x$Fe$_2$As$_2$: evidence for multiband strong-coupling behavior}

\author{A.\,A.~Golubov$^1$, O.\,V.~Dolgov$^2$, A.\,V.~Boris$^2$, A.~Charnukha$^2$,
D.\,L.~Sun$^2$, C.\,T.~Lin$^2$,
A.\,F.~Shevchun$^3$, A.\,V.~Korobenko$^{3,4}$, M.\,R.~Trunin$^{3,4}$, and V.\,N.~Zverev$^{3,4}$}
\rauthor{A.\,A.~Golubov, O.\,V.~Dolgov, A.\,V.~Boris, A.~Charnukha,
D.\,L.~Sun, C.\,T.~Lin, A.\,F.~Shevchun, A.\,V.~Korobenko, M.\,R.~Trunin, and V.\,N.~Zverev}
\sodauthor{A.\,A.~Golubov, O.\,V.~Dolgov, A.\,V.~Boris, A.~Charnukha,
D.\,L.~Sun, C.\,T.~Lin, A.\,F.~Shevchun, A.\,V.~Korobenko, M.\,R.~Trunin, and V.\,N.~Zverev}

\address{
$^1$ Faculty of Science and Technology and MESA+ Institute for Nanotechnology, University of Twente, Enschede, The Netherlands \\
$^2$ Max-Planck-Institut fur Festkorperforschung, Heisenbergstrasse 1, D-70569 Stuttgart, Germany \\
$^3$ Institute of Solid State Physics, Chernogolovka, Moscow district, Russia \\
$^4$ Moscow Institute of Physics and Technology, Dolgoprudny, Moscow Region, Russia}

\abstract{We present theoretical analysis of the normal state resistivity in multiband
superconductors in the framework of Eliashberg theory. The results are compared with measurements
of the temperature dependence of normal state resistivity of high-purity Ba$_{0.68}$K$_{0.32}$Fe$_{2}$As$_{2}$ single crystals
with the highest reported transition temperature $T_c$ = 38.5 K. The experimental data demonstrate strong deviations
from the Bloch-Gr\"{u}neisen behavior, namely the tendency to saturation of the resistivity at high temperatures.
The observed behavior of the resistivity is explained within the two band scenario when the first band
is strongly coupled and relatively clean, while the second band is weakly coupled and is
characterized by much stronger impurity scattering.}

\maketitle

\textbf{Introduction.} It is widely known that many disordered alloys exhibit resistivity
saturation \cite{Mooij,Fisk}.
In order to
describe these data, phenomenological model was suggested which assumed the
existence of an effective shunt with large temperature-independent
resistance. This model assumes the existence of two parallel conductivity
channels, i.e. two groups of carriers having different
scattering parameters:

\begin{equation}
\rho ^{-1}=\rho _{0}^{-1}+\rho _{sh}^{-1},  \label{Eq1}
\end{equation}%
where $\rho _{0}$ is the resistivity of the first group of carriers which is
characterized by weak scattering on static defects and by large slope of temperature
dependence of resistivity (strong temperature-dependent scattering) and $\rho _{sh}
$ is the resistivity of the second group (shunt) characterized by strong
temperature-independent scattering of carriers. This simple approach provides good
fit to experimental data \cite{Mooij,Fisk,Gurvich} but it was never justified on
physical grounds (see the discussion in review \cite{Gantmakher}). Resistivity
saturation was also observed in previous studies of V$_{3}$Si compounds \cite
{Milewits,Nefyodov} and an empirical explanation was suggested in \cite{Milewits} to
explain this effect, however no physical basis for such approach was
provided.

The newly discovered iron pnictide superconductors
\cite{Hosono} present an unusual case of multiband superconductivity.
Currently, there is accumulating evidence in favor of multiband effects and
pairing mechanism due to exchange of magnetic fluctuations connecting
different sets of Fermi surfaces \cite{Mazin1,Kuroki,Mazin2,Johnston}. Apart
of superconducting behavior, normal state properties of pnictides also
attract a lot of interest. Recent measurements of normal state resistivity
and Hall effect revealed a number of anomalous features. While the
resistivity of Co-doped BaFe$_{2}$As$_{2}$ compounds follows standard
Bloch-Gr\"{u}neisen low above $T_{c}$ \cite{Co,Co1}, in K- and Ru-doped BaFe$%
_{2}$As$_{2}$, as well as in LiFeAs and SrPt$_{2}$As$_{2}$ compounds the
resistivity exhibit a tendency to saturation \cite{K,Zverev,Li,Ru,Pt}. In
addition, Hall coefficient in these materials is temperature-dependent.

In the present work we will argue that the effective shunt model can be
derived for hole-doped pnictides and it explains the resistivity saturation
observed in these compounds. The model is based on our knowledge of
electronic structure of pnictides. In Ref.~\cite{Boris}, a microscopic
calculation of the specific heat in the framework of a four band Eliashberg
spin-fluctuation model was performed. These four bands correspond to two
electron pockets and two hole pockets. It was further shown in \cite
{Boris} that two effective bands can be constructed by combining two
electronic pockets and inner hole pocket (around $\Gamma $-point) into one
band, while the remaining outer hole pocket forms the second band. Important
result is that the second band is characterized by much weaker
interaction of carriers with intermediate bosons (presumably spin fluctuations)
than the first one, since nesting conditions
are not fulfilled for the outer hole pocket. Further, the authors
of Ref. \cite{Co1} reached important conclusion that
relaxation rates of the holes in
Ba(Fe$_{1-x}$Co$_{x}$)$_{2}$As$_{2}$
are much higher than relaxation rates of the electrons,
as follows from their analysis of Hall effect in this compound. As a result, the
outer hole pocket is characterized by weak interaction between carriers and
strong temperature-independent scattering and thus provides the physical
realization of effective shunting resistor in the model \cite%
{Mooij,Fisk,Gurvich,Gantmakher}. 


In this work we perform theoretical analysis of a normal state resistivity
and compare it with the data for \textit{dc} and microwave resistivity
measurements in Ba$_{0.68}$K$_{0.32}$Fe$_{2}$As$_{2}$ single crystals with
highest available critical temperature $T_{c}$ = 38.5 K. The proposed
multiband scenario is also consistent with our Hall-effect measurements on
the same Ba$_{0.68}$K$_{0.32} $Fe$_{2}$As$_{2}$ single crystals. Our model
also provides explanation of the difference between resistivities in
K- and Co-doped BaFe$_{2}$As$_{2}$ in terms of stronger impurity scattering
within electronic pockets in Co-doped compounds.

\textbf{Single band case.} The DC resistivity in the single band model is
determined by the expression \cite{Grimwall,Allen}

\begin{equation}
\rho _{DC}(T)=\left[ \frac{\epsilon_0 \omega _{pl\text{ }}^{2}}{W}\right]
^{-1},  \label{rho1}
\end{equation}%
where $\omega _{pl\text{ }}$ is a bare (band) plasma frequency, and the
kernel $W(T)$ is determined by the impurity scattering rate $\gamma ^{imp}$
and the Bloch-Gr\"{u}neisen integral of the \textit{transport} Eliashberg
function $\alpha _{tr\text{ }}^{\text{ }2}(\Omega )F(\Omega )$

\begin{equation}
W=\gamma ^{imp}+\frac{\pi }{\beta T}\int_{0}^{\infty }d\Omega \frac{\Omega }{%
\sinh ^{2}(\Omega /2 \beta T)}\alpha _{tr\text{ }}^{\text{ }2}(\Omega
)F(\Omega ).  \label{Wrho1}
\end{equation}%
where $\beta=k_B/\hbar$. To simplify notations, below we will skip the
subscript '$tr$' in the transport Eliashberg function and coupling constants.

\textbf{Two band case.} In the two-band case the above expression for the
resistivity can be straightforwardly extended by adding the conductivities
of both bands
\begin{equation}
\rho _{DC}(T)=\left[ \frac{\epsilon_0 \omega _{pl\text{ }1}^{2}}{W_{1}}+%
\frac{\epsilon_0 \omega _{pl\text{ }2}^{2}}{W_{2}}\right] ^{-1},
\label{rho2}
\end{equation}

\begin{equation}
W_{i}=\gamma _{i}^{imp}+\frac{\pi }{\beta T}\int_{0}^{\infty }d\Omega \frac{%
\Omega }{\sinh ^{2}(\Omega /2 \beta T)}\alpha _{i}^{\text{ }2}(\Omega
)F_{i}(\Omega ),  \label{Wrho2}
\end{equation}%
where $i=1,2$. In the above expression the scattering parameters are defined
as $\gamma _{1}^{imp}=\gamma _{11}^{imp}+\gamma _{12}^{imp}$ and $\gamma
_{2}^{imp}=\gamma _{21}^{imp}+\gamma _{22}^{imp}$, where $\gamma
_{11}^{imp},\gamma _{22}^{imp}$ and $\gamma _{12}^{imp},\gamma _{21}^{imp}$
are, respectively, \emph{intraband} and \emph{interband} scattering rates.

At high temperatures, $T\gtrsim \tilde{\Omega}/5$, where $%
\tilde{\Omega}$ is the characteristic energy of the Eliashberg function, the
expression (\ref{Wrho1}) has a form

\[
W=\gamma ^{imp}+2\pi \lambda \beta T,
\]%
where the transport coupling constant $\lambda $ is defined by the standard
relation
\begin{equation}
\lambda =2\int_{0}^{\infty }d\Omega \frac{\alpha ^{2}(\Omega )F(\Omega )}{%
\Omega }.
\end{equation}%
In this regime

\begin{equation}
\rho _{DC}(T)=\left[ \frac{\epsilon_0 \omega _{pl\text{ }1}^{2}}{\gamma
_{1}^{imp}+2\pi \lambda _{1} \beta T}+\frac{\epsilon_0 \omega _{pl\text{ }%
2}^{2}}{\gamma _{2}^{imp}+2\pi \lambda _{2} \beta T}\right] ^{-1},
\label{rho}
\end{equation}%
where the effective coupling constants are defined as $\lambda _{1}=\lambda
_{11}+\lambda _{12}$ and $\lambda _{2}=\lambda _{21}+\lambda _{22}$ and
\begin{equation}
\lambda _{ij}=2\int_{0}^{\infty }d\Omega \frac{\alpha _{ij}^{\text{ }%
2}(\Omega )F_{ij}(\Omega )}{\Omega }.
\end{equation}

In the two band model the saturation of the resistivity has the following
explanation.
First, there exists large disparity between the coupling constants in Ba$_{1-x}$K$_x$Fe$_2$As$_2$:
$\lambda _{1}\gg \lambda _{2}$. This fact follows from the
quantitative analysis of thermodynamic data presented in Ref. \cite{Boris}.
Second, Hall effect data presented in Ref. \cite{Co1} suggest that
$\gamma _{2}^{imp}\gg \gamma _{1}^{imp}$. Though calculation of these scattering rates is
beyond the framework of the present paper, we consider this relation between
the scattering parameters as a reasonable assumption
in the two-band model for Ba$_{1-x}$K$_x$Fe$_2$As$_2$.
As a result, combining the conditions $\lambda _{1}\gg \lambda _{2}$ and
$\gamma _{2}^{imp}\gg \gamma _{1}^{imp}$ we arrive the expression
\begin{equation}
\rho _{DC}(T)=\left[ \frac{\epsilon_0 \omega _{pl\text{ }1}^{2}}{\gamma
_{1}^{imp}+2\pi \lambda _{1} \beta T}+\frac{\epsilon_0 \omega _{pl\text{ }%
2}^{2}}{\gamma _{2}^{imp}}\right] ^{-1},
\end{equation}%
which is valid in a broad temperature range $T < \gamma _{2}^{imp}/2\pi \lambda _{2} \beta  $.
As follows from the above expression, the resistivity saturates at $%
T>\gamma _{2}^{imp}/2\pi \lambda _{1} \beta$ up to the value%
\[
\rho _{DC}(T)=\frac{\gamma _{2}^{imp}}{\epsilon_0 \omega _{pl\text{ }2}^{2}}%
.
\]%
This saturation manifests itself as the crossover from the standard Bloch-Gr%
\"{u}neisen temperature dependence of the resistivity to the behavior
characterized by the downward curvature of the $\rho (T)$ curve, with the
crossover temperature $T^{\ast }\simeq \gamma _{2}^{imp}/2\pi \lambda _{1}
\beta $.

Thus, we have shown that at high temperatures the conductivity of the second band shunts the
conductivity of the first one, leading to the saturation
of the total resistance (see Eq.1). Below we apply this model to
describe the experimental data for the resistivity of Ba$_{1-x}$K$_{x}$Fe$_{2}$As$_{2}$ compound.

\textbf{Comparison with experiment and discussion.} The transport properties
of hole-doped Ba$_{1-x}$K$_{x}$Fe$_{2}$As$_{2}$ single crystals in normal
and superconducting state were studied by two different techniques: by
measuring the microwave surface impedance and by measuring $dc $
resistivity. The measurements were carried out on single crystals of $\text{%
Ba}_{1-x}\text{K}_{x}\text{Fe}_{2}\text{As}_{2}$ (BKFA) with $x=0.32$ and
superconducting $T_{\mathrm{c}}=38.5\ \text{K}$. Specific heat measurements
on the same samples confirm there high quality and the absence of secondary
electronic phases ~\cite{Boris}.

\begin{figure}[t]
\includegraphics[width=8cm]{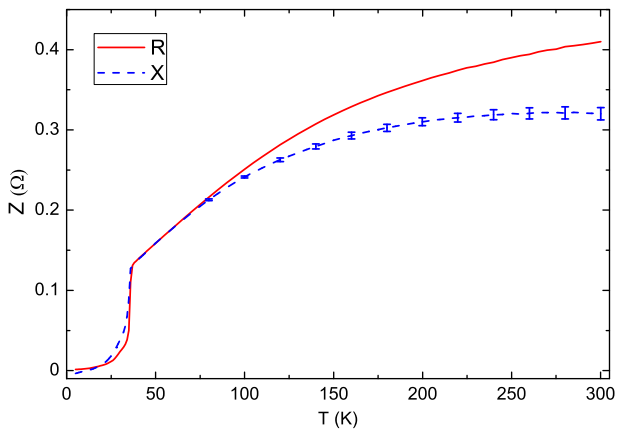}
\caption{Fig.\protect\ref{fig:fig1}. Surface resistance (R) and reactance (X) in
the conducting layers of a Ba$_{1-x}$K$_{x}$Fe$_{2}$As$_{2}$ single
crystal at frequency 9.42 GHz. Sample size is 1.65x0.8x0.06 mm$^{3}$.}
\label{fig:fig1}
\end{figure}

The temperature dependence of the surface impedance of the Ba$_{1-x}$K$_{x}$%
Fe$_{2}$As$_{2}$ single crystals were measured by the "hot finger" technique
at frequency 9.42 GHz. The experimental setup involved a cylindrical cavity
niobium resonator operating at the mode $H_{011}$. The walls of the
resonator are cooled down with liquid helium and are in the superconducting
state. The crystal was placed at the end of the sapphire rod in a uniform
high-frequency magnetic field, so that microwave currents flow along the
superconducting layers of the crystal. The temperature of the rod and the
sample can be varied in the range from 5 to 300 K.

In the temperature range $40~K<T<80~K$ one can see normal skin effect: the
real (surface resistance $R(T)$) and imaginary (reactance $X(T)$) parts of
the surface impedance are equal, $R(T)=X(T)$ (see Fig.1). At $T>80~K$ the
reactance $X(T)$ becomes less than $R(T)$, which is most likely due to
thermal expansion of the crystal \cite{Trunin}. Fig.2 shows temperature
dependence of the resistivity found from the expression $\rho
(T)=2R^{2}(T)/\omega \mu _{0}$ valid for a homogeneous conductor.

The $dc$ resistivity measurements were carried out in the temperature range $%
4.2~K<T<280~K$. The samples were thin plates with 1.8x0.8x0.02 mm$^{3}$
characteristic size. The largest surface of the plate was oriented along the
conducting layers. The sample resistance was measured using a four-probe
technique by a lock-in detector at 20 Hz alternating current. The contacts
were prepared on the sample surface with conducting silver paste. The
in-plane resistivity component $\rho (T)$ is characterized by anomalous
T-dependence: $\rho (T)$ curve is convex with the tendency to saturate at
high temperature, i.e. similar to the $R(T)$ dependence measured at
microwave frequency.

\begin{figure}[b]
\includegraphics[width=8cm]{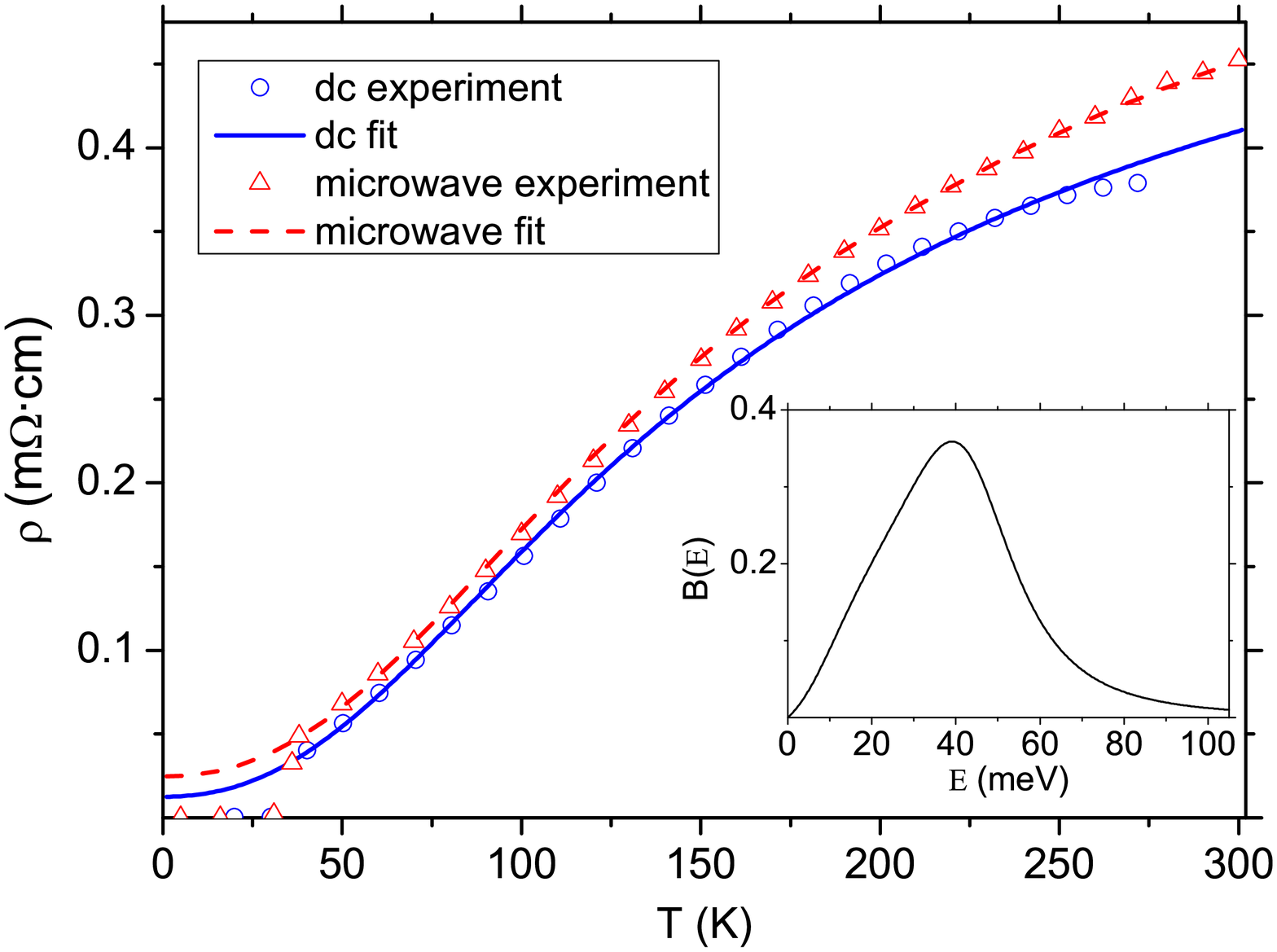}
\caption{Fig.\protect\ref{fig:fig2}. Theoretical fits to resistivity of Ba$%
_{1-x}$K$_{x}$Fe$_{2}$As$_{2}$ single crystals determined from \textit{dc}
and microwave measurements. The fitting parameters are given in the text.}
\label{fig:fig2}
\end{figure}

The results of comparison between theory and experiment are shown in Fig.2.
Important parameters in the present model are impurity scattering rates in
each band. As is seen from the figure, the results of measurements are
consistent with the scenario when the first (strongly coupled) band has much
smaller scattering rate than the second one. Indeed, in accordance with
theoretical picture described above, resistivity saturation occurs when
two conditions, $\lambda _{1}\gg \lambda _{2}$ and $\gamma _{2}^{imp}\gg \gamma _{1}^{imp}$,
are fulfilled.

In the framework of
spin-fluctuation paring mechanism \cite{Mazin1,Kuroki,Mazin2}, nesting
between electron and hole pockets is important for superconductivity. Such
nesting conditions are realized for inner hole pocket and electronic
pockets, but are not fulfilled for the outer hole pocket. Therefore, as was
pointed out above, the second band is characterized by much weaker pairing
interaction than the first one. Our fitting parameters are in qualitative
agreement with this scenario, namely $\lambda _{1}$=1.8, $\lambda _{2}\simeq0
$. The scattering rates obtained from the fitting are: $\gamma _{1}^{imp}$=3
meV, $\gamma _{2}^{imp}$=67 meV (for the dc resistivity fit) and $\gamma
_{1}^{imp}$=6 meV, $\gamma _{2}^{imp}$ = 78 meV (for the microwave
resistivity fit). The differences between the dc and microwave resistivity
fits can be attributed to the fact that microwave impedance measurement is
surface sensitive technique. Plasma frequencies were chosen as $\omega _{pl1}
$ = 1.33 eV and $\omega _{pl2}$ = 0.87 eV and are consistent with optical
measurements \cite{Charnukha} were total plasma frequency of 1.6 eV was
determined. The transport Eliashberg function was chosen in the form $\alpha
_{i}^{2}(\Omega )F(\Omega )=\lambda _{i} B(\Omega)$, where the function $%
B(\Omega)$ is shown in the inset in Fig.2. With the above parameters, the
crossover temperature $T^{\ast }\simeq \gamma _{2}^{imp}/2\pi \lambda _{1}
\beta \sim $ 100 K.
The estimated value of the Fermi energy $E_F$ is equal to 0.6 eV at $T=300$ K.
Therefore, up to the highest temperatures the Ioffe-Regel limit in energy units
$E_F\sim(\gamma_i^{imp}+2\pi \lambda_i \beta T)$ is
not reached for both bands and localization effects (neglected in our model) are not important.
Note also that despite a number of free parameters, the
present model provides rather unique fit, since the crossover only occurs in
narrow parameter range, corresponding to strong, about an order of
magnitude, disparity of the scattering rates and the coupling strengths in
both bands.

\begin{figure}[bh]
\includegraphics[width=8cm]{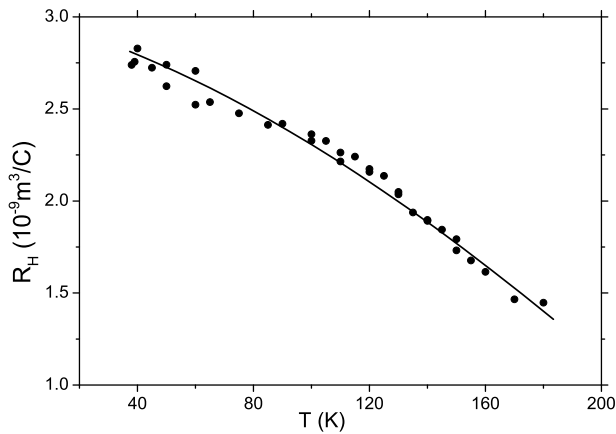}
\caption{Fig.\protect\ref{fig:fig3}. The temperature dependence of the Hall
constant in a Ba$_{1-x}$K$_{x}$Fe$_{2}$As$_{2}$ single crystal. Solid line
is a guide to an eye.}
\label{fig:fig3}
\end{figure}

The multiband scenario, applied above to explain the resistivity anomaly at
high temperatures, is also consistent with our results of Hall effect
measurements on the same set of samples shown in Fig.3. One can see that
Hall constant $R_{H}$ decreases with increasing temperature. The temperature
dependent $R_{H}$ value was also observed recently \cite{Co,Co1} in Co-doped
and K-doped compounds. Though detailed quantitative description of the
behavior of the Hall constant is beyond the framework of the present work,
qualitative discussion was given in Ref. \cite{Co1}. As is argued in \cite%
{Co1}, the temperature dependence of $R_{H}$ is naturally explained by the
mobility changes, because in multi-band conductors $R_{H}$ value is the
function of both concentration and mobility. 

The difference in resistivity behavior between K- and Co- doped samples has
natural explanation in the framework of our model. The transition
temperature in Co- doped pnictides is reduced, which can be attributed to
weaker coupling constant $\lambda _{1}$ in the first band. Within our
scenario, that means that
the crossover temperature $T^{\ast }\simeq \gamma_{2}^{imp}/2\pi \lambda
_{1} \beta$ is higher than in the K-doped case considered above. As a
result, the regime of the resistivity saturation in Co-doped pnictides
should shift to higher temperatures.

In conclusion, we have presented two-band Eliashberg model for the
normal-state resistivity in iron-pnictides, which naturally explains the
observed tendency to resistivity saturation and temperature-dependent Hall
coefficient in K- and Ru-doped BaFe$_{2}$As$_{2}$, as well as in LiFeAs and
SrPt$_{2}$As$_{2}$ compounds. The results are in a good agreement with
\textit{dc} and microwave resistivity measurements in Ba$_{0.68}$K$_{0.32}$Fe%
$_{2}$As$_{2}$ single crystals with $T_{c}$ = 38.5 K. This analysis reveals
strong disparity of relaxation rates and strongly different coupling
constants in different bands in these materials.

\textbf{Acknowledgements}. We acknowledge useful discussions with V.F.
Gantmakher, R.K. Kremer, I.I. Mazin and A.N. Yaresko. This work was
supported by the Grant 2009-1.5-508-008-043 within the Federal program
funded by Russian Ministry of Education and Science and by RFBR grants
Nos. 09-02-01224 and 11-02-12071.

\end{document}